\documentclass[aps,prl,twocolumn]{revtex4}

\usepackage{epsf}
\usepackage{bm}

\begin{document}

\title{Universal features of the optical properties of ultrathin plasmonic films}

\author{Igor V.~Bondarev}\email[E-mail: ]{ibondarev@nccu.edu}
\affiliation{Math \& Physics Department, North Carolina Central University, Durham, NC 27707, USA}

\author{Vladimir M.~Shalaev}
\affiliation{School of Electrical \& Computer Engineering and Birck Nanotechnology Center, Purdue University, West Lafayette, IN 47907, USA}

\begin{abstract}
We study theoretically confinement related effects in the optical response of thin plasmonic films of controlled variable thickness. While being constant for relatively thick films, the plasma frequency is shown to acquire spatial dispersion typical of two-dimensional materials such as graphene, gradually shifting to the red with the film thickness reduction. The dissipative loss, while decreasing at any fixed frequency, gradually goes up at the plasma frequency as it shifts to the red with the film thickness reduced. These features offer a controllable way to tune spatial dispersion and related optical properties of plasmonic films and metasurfaces on demand, by precisely controlling their thickness, material composition, and by choosing deposition substrates and coating layers appropriately.
\end{abstract}

\maketitle


Current development of nanofabrication techniques makes it possible to design advanced plasmonic nanomaterials with optical properties on-demand~\cite{Liz-Marzan,Stokes}. One type of such advanced nanomaterials are optical metasurfaces (see, e.g., Ref.~\cite{Shalaevgroup13} for review). Metasurfaces are often based on thin quasi-two-dimensional (2D) plasmonic films, which enable new physics and phenomena that are distinctly different from those observed for their 3D counterparts~\cite{GarciaAbajo14,Shalaevgroup16,Shalaevgroup17,Mak,SlavaPopov,Koppens15,Basov1,Basov2,Koppens14}. Nowadays, a careful control of the geometry, structural dimensions, and material composition allows one to produce thin and ultrathin metasurfaces for applications in optoelectronics, ultrafast information technologies, microscopy, imaging, and sensing as well as for probing the fundamentals of the light-matter interactions at the nanoscale~\cite{Rodrigo,Atwater,Halas,West,Koppens11}. A key to realizing these applications is the ability to fabricate metallic films of precisely controlled thickness down to a few monolayers, which also exhibit desired optical properties~\cite{Shalaevgroup17}. As the film thickness decreases, the strong electron confinement could lead to new confinement related effects~\cite{GarciaAbajo14}, which require theory development to understand their role in the light-matter interactions and optical response of thin and ultrathin plasmonic films.

We develop a quasiclassical theory for the electron confinement effects and their manifestation in the optical response of thin plasmonic films of variable thickness. We start with the Coulomb interaction potential in the confined planar geometry to obtain the equations of motion and the conditions for the in-plane collective electron motion. The plasma frequency thus obtained, while being constant for relatively thick films, acquires spatial dispersion typical of 2D materials and gradually shifts to the red as the film thickness decreases. The complex-valued dynamical dielectric response function shows the gradual red shift of the resonance frequency, accordingly, with the dissipative loss decreasing at any fixed frequency and gradually going up at the plasma frequency as it shifts to the red with the film thickness reduction. These are the universal features peculiar to all plasmonic thin films, which can be controlled not only by varying the thickness and material composition of the film, but also by choosing deposition substrates and coating layers appropriately.

\begin{figure}[t]
\epsfxsize=8.65cm\centering{\epsfbox{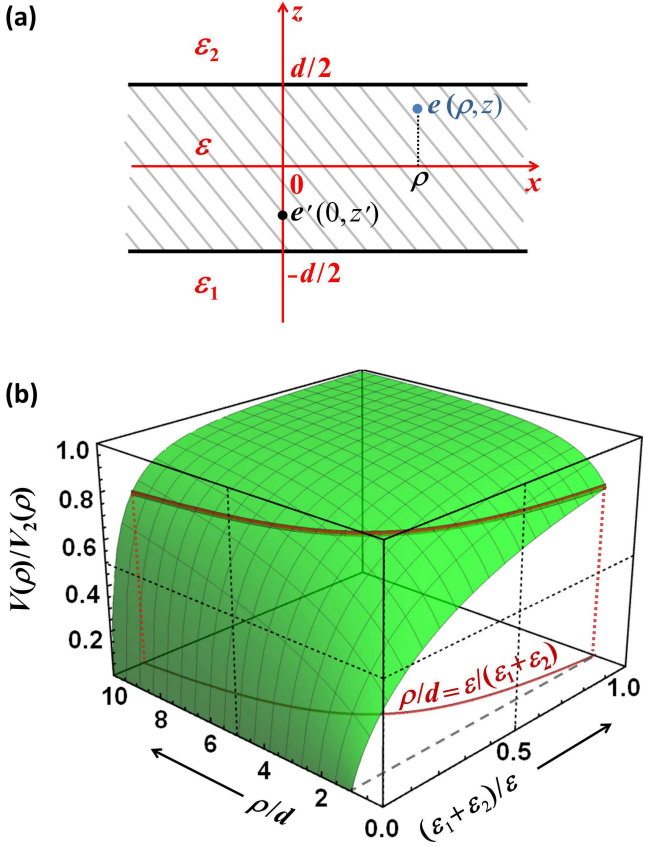}}\caption{(Color online) (a,b) Schematic to show the geometry notations and the normalized electrostatic (Keldysh) potential for the Coulomb interaction of the two point charges $e$ and $e^\prime$ confined in a planar thin film of finite thickness.}\label{fig1}
\end{figure}

The Coulomb interaction in thin films (see Fig.~\ref{fig1}) increases strongly with decreasing film thickness if the film dielectric constant is much larger than those of the film surroundings~\cite{Keldysh79}. In general, any kind of spatial confinement results in the increase of the Coulomb interaction between charges inside the confined structure as the characteristic confinement size decreases, provided that the environment has a lower dielectric constant than that of the confined structure. This is because the field produced by the confined charges outside of the confined structure begins to play a perceptible role with the confinement size reduction. If the surrounding medium dielectric constant is much less than that of the confined structure, then the increased 'outside' contribution to the Coulomb energy makes the Coulomb interaction between the charges confined stronger than that in a homogeneous medium with the confined structure dielectric constant. Specific examples to confirm this fact, theoretical and experimental ones, can be found in the literature both for quasi-1D and for quasi-2D confined geometries~\cite{Francois02,Louie09,TonyHeinz14}.

For the thin film of thickness $d$ with the background dielectric constant $\varepsilon$, surrounded by media with the dielectric constants $\varepsilon_1$ and $\varepsilon_2$ as shown in Fig.~\ref{fig1}~(a), the Coulomb potential between the two confined charges $e$ and $e^\prime$ loses its $z$-coordinate dependence to turn into a pure in-plane 2D potential when $\varepsilon_{1,2}\!\ll\!\varepsilon$ and the in-plane inter-charge distance $\rho\!\gg\!d$. The potential takes the form (first reported by L.V.Keldysh~\cite{Keldysh79})
\begin{equation}
V(\rho)=\frac{\pi ee^\prime}{\varepsilon d}\left[H_0\!\left(\!\frac{\varepsilon_1+\varepsilon_2}{\varepsilon}\frac{\rho}{d}\right)
-N_0\!\left(\!\frac{\varepsilon_1+\varepsilon_2}{\varepsilon}\frac{\rho}{d}\right)\right],
\label{Keldysh1}
\end{equation}
where $N_0(x)$ and $H_0(x)$ are the Neumann and Struve functions, respectively. This (Keldysh) potential can be further expanded to give two different asymptotic expressions as follows
\begin{equation}
V(\rho)\approx V_1(\rho)=\frac{2ee^\prime}{\varepsilon d}\left[\ln\!\left(\!\frac{2\varepsilon}{\varepsilon_1+\varepsilon_2}\frac{d}{\rho}\right)-C\right],
\label{Keldysh2}
\end{equation}
($C\!\approx\!0.577$ is the Euler constant) if $\varepsilon/(\varepsilon_1+\varepsilon_2)\!\gg\!\rho/d\!\gg\!1$, and
\begin{equation}
V(\rho)\approx V_2(\rho)=\frac{2ee^\prime}{(\varepsilon_1+\varepsilon_2)\rho}\,,
\label{Keldysh3}
\end{equation}
if $\rho/d\!\gg\!\varepsilon/(\varepsilon_1+\varepsilon_2)\!\gg\!1$. This latter expression includes no parameters to represent the thin film itself, and it shows no screening at all for the film in air ($\varepsilon_1\!=\!\varepsilon_2\!=\!1$), which is quite an interesting result.

Figure~\ref{fig1}~(b) presents the normalized Keldysh potential [the ratio $V(\rho)/V_2(\rho)$ with $V(\rho)$ and $V_2(\rho)$ given by Eqs.~(\ref{Keldysh1}) and (\ref{Keldysh3}), respectively] as a function of $\rho/d$ and the relative dielectric constant $(\varepsilon_1+\varepsilon_2)/\varepsilon$. The potential is seen to vary drastically in the domain where $\varepsilon\!\gg\!\varepsilon_1+\varepsilon_2$ and $d\!\ll\!\rho$, which is just the parameter range for thin plasmonic films~\cite{GarciaAbajo14,Shalaevgroup17,Shalaevgroup16,Shalaevgroup13}. The drastic change of the Coulomb interaction potential in this domain comes from Eq.~(\ref{Keldysh2}), which varies much faster than any Coulomb type ($\sim\!1/\rho$) potential does --- a solely confinement related effect associated with the strong spatial dispersion of the thin-film dielectric response function, the dielectric permittivity.

One can easily evaluate the plasma frequency spatial dispersion in finite-thickness plasmonic films ($\varepsilon$) sandwiched [Fig.~\ref{fig1}~(a)] between dielectric substrates ($\varepsilon_1$) and superstrates ($\varepsilon_2$). Using the in-plane (2D) Fourier expansion of the Keldysh potential in Eq.~(\ref{Keldysh1}), the Coulomb potential energy of the quasi-free outer-shell electron located at the point $\bm{\rho}_j\!=\!(\rho_j,\varphi_j)$ of the lattice site $j$, which interacts with the other electrons of the finite-thickness thin film --- all immersed in the positive background of the film material ("jellium" model~\cite{Pines}), takes the form
\begin{equation}
V(\bm{\rho}_j)=\frac{4\pi e^2}{\varepsilon L^2}
\sum_{l,\mathbf{k}}~^{\!\!\!\textstyle^\prime}\frac{\exp\left(i\mathbf{k}\!\cdot\!{\bm\rho}_{jl}\right)}{k[kd+(\varepsilon_1+\varepsilon_2)/\varepsilon]}\,.
\label{KeldyshFourier}
\end{equation}
Here, $\!\int\!d\mathbf{k}\exp[i\mathbf{k}\cdot(\bm\rho-\bm\rho^\prime)]/(2\pi)^2\!=\delta(\bm\rho-\bm\rho^\prime)$ is used as the normalization condition for the Fourier expansion basis function set, with $\mathbf{k}$ representing the in-plane electron quasi-momentum ($k_{x,y}\!=\!2\pi n_{x,y}/L$ with $n_{x,y}\!=\!0,\pm1,...$; $L\!\gg\!d$ stands for the square-sized film length), $k\!=\!|\mathbf{k}|$, and ${\bm\rho}_{jl}\!=\!{\bm\rho}_j-{\bm\rho}_l$. The summation sign is primed to indicate that the terms with $l\!=\!j$ and $\mathbf{k}\!=\!0$ associated with the electron self-interaction and with the all-together electron displacement, respectively, must be dropped.

Using Eq.~(\ref{KeldyshFourier}) along with the electron kinetic energy $K\!=\!\sum_lm^\ast\dot{\bm\rho}_l^{\,2}/2$, where $m^\ast$ is the electron effective mass, one arrives at the individual electron equations of motion
\begin{equation}
\ddot{\bm\rho}_j=-i\frac{4\pi e^2}{\varepsilon m^\ast L^2}\sum_{l,\mathbf{k}}~^{\!\!\!\textstyle^\prime}
\frac{\textbf{k}\exp\left(i\mathbf{k}\!\cdot\!\bm\rho_{jl}\right)}{k[kd+(\varepsilon_1+\varepsilon_2)/\varepsilon]}\,.
\label{acceleration}
\end{equation}
To obtain the equations of motion for the density of electrons, we introduce the local surface electron density
\[
n(\bm\rho)=\sum_l\delta(\bm\rho-\bm\rho_l)=\sum_\mathbf{k}n_\mathbf{k}\exp\left(i\mathbf{k}\!\cdot\!\bm\rho\right)\nonumber\\
\]
with the Fourier components
\begin{equation}
n_\mathbf{k}=\frac{1}{L^2}\sum_l\exp\left(-i\mathbf{k}\!\cdot\!\bm\rho_l\right),\;\;\;n_{\mathbf{k}=0}=N_{2D}\hskip0.5cm
\label{N2D}
\end{equation}
($N_{2D}$ being the equilibrium \emph{surface} electron density), and use Eq.~(\ref{acceleration}) to get the second time derivative in the form
\begin{eqnarray}
\ddot{n}_\mathbf{k}=-\frac{4\pi e^2}{\varepsilon m^\ast}\sum_{\mathbf{q}}~^{\!\!\!\textstyle^\prime}
\frac{\left(\textbf{k}\!\cdot\!\mathbf{q}\right)n_\mathbf{q}n_{\mathbf{k}-\mathbf{q}}}{q[qd+(\varepsilon_1+\varepsilon_2)/\varepsilon]}\nonumber\\
-\frac{1}{L^2}\sum_l\left(\textbf{k}\!\cdot\!\dot{\bm\rho}_l\right)^2\exp\left(-i\mathbf{k}\!\cdot\!\bm\rho_l\right).\hskip0.1cm\nonumber
\end{eqnarray}
This can now be simplified in the random phase approximation (RPA) by dropping alternating-sign terms ($\mathbf{q}\!\ne\!\mathbf{k}$) in the sum over $\mathbf{q}$~\cite{Pines}, to obtain after using $N_{2D}$ of Eq.~(\ref{N2D})
\begin{eqnarray}
\ddot{n}_\mathbf{k}+\omega_p^2\,n_\mathbf{k}=-\frac{1}{L^2}\sum_l\left(\textbf{k}\!\cdot\!\dot{\bm\rho}_l\right)^2\exp\left(-i\mathbf{k}\!\cdot\!\bm\rho_l\right)
\label{maineqnrpa}
\end{eqnarray}
with
\begin{equation}
\omega_p=\omega_p(k)=\sqrt{\frac{4\pi e^2kN_{2D}}{\varepsilon m^\ast[kd+(\varepsilon_1+\varepsilon_2)/\varepsilon]}}\,.
\label{omegapofk}
\end{equation}
Equation~(\ref{maineqnrpa}) is seen to turn into the oscillator equation provided $\mathbf{k}^2\!\ll\!\mathbf{k}_c^2\!=\!\omega_p^2/v_0^2$ with $v_0$ given by $m^\ast v_0^2/2\!=\!E_F$, or $k_BT$ for the degenerate and non-degenerate electron gas system, respectively~\cite{Pines}. When electron wave vectors are much less than the cut-off vector $\mathbf{k}_c$, the right hand side of Eq.~(\ref{maineqnrpa}) becomes much less than $\mathbf{k}_c^2v_0^2\,n_{\mathbf{k}}\!=\omega_p^2\,n_{\mathbf{k}}$, yielding the thin-film electron density coherent oscillations with the plasma frequency featuring the spatial dispersion given by Eq.~(\ref{omegapofk}), as opposed to bulk (3D) plasmonic materials in which case the plasma frequency is the $k$-independent constant
\begin{equation}
\omega_p^{3D}=\sqrt{\frac{4\pi e^2N_{3D}}{\varepsilon m^\ast}}
\label{bulkplasmafreq}
\end{equation}
with $N_{3D}$ representing the \emph{volumetric} electron density.

For thin enough plasmonic films, one has $N_{3D}d=\!N_{2D}$, so that Eq.~(\ref{omegapofk}) can be written as
\begin{equation}
\omega_p=\omega_p(k)=\frac{\omega_p^{3D}}{\sqrt{1+(\varepsilon_1+\varepsilon_2)/\varepsilon kd}}\,.
\label{omegap}
\end{equation}
If $(\varepsilon_1+\varepsilon_2)/\varepsilon kd\ll\!1$ (relatively thick film), then $\omega_p\!=\omega_p^{3D}$ of Eq.~(\ref{bulkplasmafreq}), whereas one has
\begin{equation}
\omega_p=\omega_p^{2D}(k)=\sqrt{\frac{4\pi e^2N_{2D}k}{(\varepsilon_1+\varepsilon_2)m^\ast}}
\label{omegap2D}
\end{equation}
if $(\varepsilon_1+\varepsilon_2)/\varepsilon kd\!\gg\!1$ (ultrathin film), which agrees precisely with the plasma frequency spatial dispersion of the 2D electron gas in air (see, e.g., Ref.~\cite{Davies}), but does show the explicit dependence on bottom ($\varepsilon_1$) and top ($\varepsilon_2$) surrounding materials.

\begin{figure}[t]
\epsfxsize=8.65cm\centering{\epsfbox{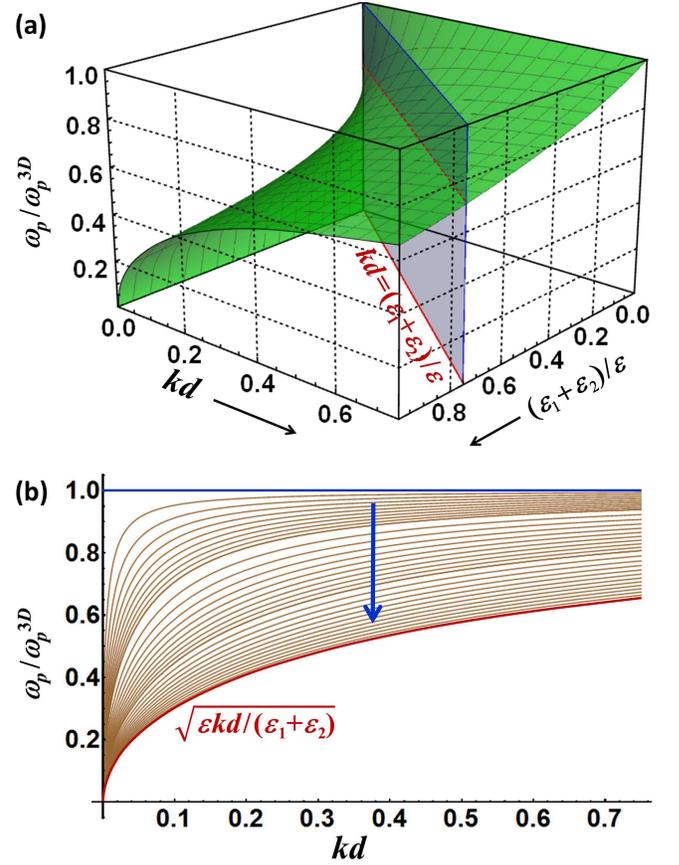}}\caption{(Color online) (a)~The ratio $\omega_p/\omega_p^{3D}$ given by Eq.~(\ref{omegap}) as a function of the dimensionless variables $kd$ and $(\varepsilon_1+\varepsilon_2)/\varepsilon$. (b)~The contour plot of the same ratio as a function of $kd$ obtained by cutting the graph in (a) with parallel vertical planes of constant $(\varepsilon_1+\varepsilon_2)/\varepsilon$. The thick vertical blue arrow shows the direction of the $(\varepsilon_1+\varepsilon_2)/\varepsilon$ increase.}
\label{fig2}
\end{figure}

The ratio $\omega_p/\omega_p^{3D}$ of Eq.~(\ref{omegap}), considered as a function of the dimensionless variables $kd$ and $(\varepsilon_1+\varepsilon_2)/\varepsilon$, represents a \emph{universal} conversion factor to relate the plasma frequency parameter in quasi-2D electron gas systems (thin finite-thickness plasmonic films~\cite{Shalaevgroup17,Shalaevgroup16}, graphene~\cite{Basov}, and related 2D materials~\cite{Mak,TonyHeinz14}) to that in bulk plasmonic materials.~The ratio is shown in Fig.~\ref{fig2}~(a). The regimes of the relatively thick and ultrathin films [Eqs.~(\ref{bulkplasmafreq}) and (\ref{omegap2D}), respectively] are separated by the vertical plane $kd\!=\!(\varepsilon_1+\varepsilon_2)/\varepsilon$. The ratio $\omega_p/\omega_p^{3D}$ is nearly constant in the domain where $kd\!\gg\!(\varepsilon_1+\varepsilon_2)/\varepsilon$, while being strongly dispersive in the domain $kd\!\ll\!(\varepsilon_1+\varepsilon_2)/\varepsilon$. In this latter case, $\omega_p$ of Eq.~(\ref{omegap}) goes down with the film thickness as $\sqrt{d}$ at all fixed $k$, which agrees with the recent plasma frequency ellipsometry measurements done on ultrathin TiN films of controlled variable thickness~\cite{Shalaevgroup17}. Figure~\ref{fig2}~(b) shows the contour plot of $\omega_p/\omega_p^{3D}$ as a function of $kd$ obtained by cutting Fig.~\ref{fig2}~(a) with parallel vertical planes of constant $(\varepsilon_1+\varepsilon_2)/\varepsilon$. We see the gradual graph profile change in the direction of the $(\varepsilon_1+\varepsilon_2)/\varepsilon$ increase (shown by the thick vertical arrow), offering a controllable way to adjust the spatial dispersion and related optical properties of plasmonic thin films and metasurfaces, in particular, not only by varying their thickness~\cite{Shalaevgroup17} and material composition~\cite{Shalaevgroup13}, but also by choosing the deposition substrates ($\varepsilon_1$) and coating layers ($\varepsilon_2$) appropriately.

\begin{figure}[t]
\epsfxsize=8.65cm\centering{\epsfbox{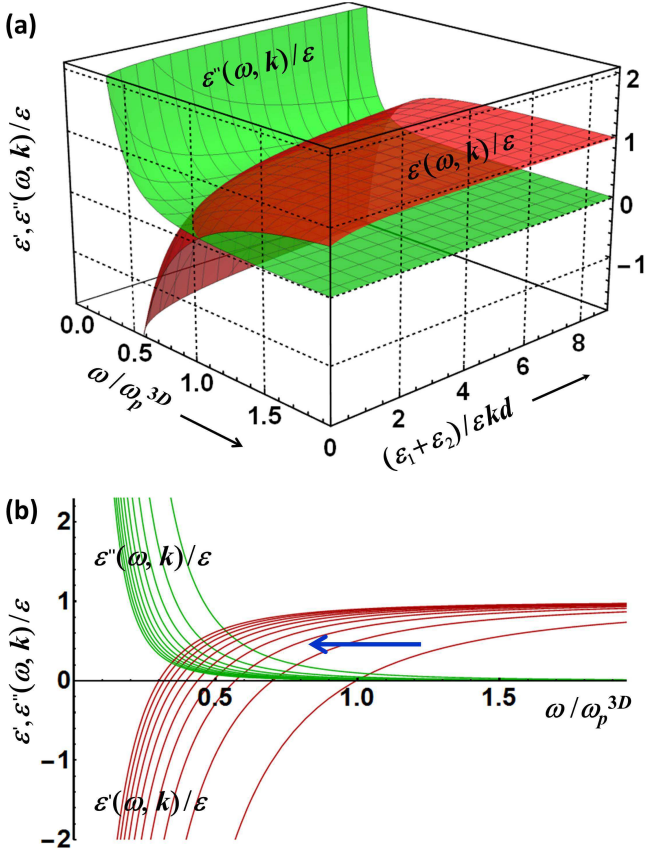}}\caption{(Color online) (a)~Real (red) and imaginary (green) parts of Eq.~(\ref{Lindhard}) as functions of the dimensionless variables $\omega/\omega_p^{3D}$ and $(\varepsilon_1+\varepsilon_2)/\varepsilon kd$. (b)~The contour plot one obtains by cutting the graph in (a) with parallel vertical planes of constant $(\varepsilon_1+\varepsilon_2)/\varepsilon kd$. The thick horizontal blue arrow shows the direction of the increase of $(\varepsilon_1+\varepsilon_2)/\varepsilon kd$.}\label{fig3}
\end{figure}

With the plasma frequency dispersion (\ref{omegap}) in hand, it is straightforward to obtain the complex-valued dynamical dielectric response function, the dielectric permittivity, for the electron gas confined in the finite-thickness ultrathin plasmonic films.~The starting point and main ingredient of the theory is the Fourier-transform of the Coulomb potential energy in Eq.~(\ref{KeldyshFourier}). With losses taken into account phenomenologically, the isotropic RPA (or Lindhard~\cite{Mahan}) low-momentum high-frequency dielectric response function $\varepsilon(k,\omega)$ (commonly referred to as the Drude response function~\cite{GarciaAbajo14}) takes the well known form
\begin{equation}
\frac{\varepsilon(k,\omega)}{\varepsilon}=1-\frac{\omega_p^{2}}{\omega(\omega+i\gamma)}\,,
\label{Lindhard}
\end{equation}
where $\gamma$ is the phenomenological inelastic electron scattering rate and $\omega_p$ is given by Eq.~(\ref{omegap}). This expression is normally used to describe the contribution of the outer-shell ($s$-band) electrons in metals~\cite{GarciaAbajo14}, with $\varepsilon$ assigned to be responsible for the positive background of the ions screened by the remaining inner-shell electrons. In many cases, however, it needs to be supplemented with an extra term (Drude-Lorentz response function~\cite{Shalaevgroup17,Shalaevgroup16}) to account for interband electronic transitions absent from Eq.~(\ref{Lindhard}).
\begin{figure}[t]
\epsfxsize=8.0cm\centering{\epsfbox{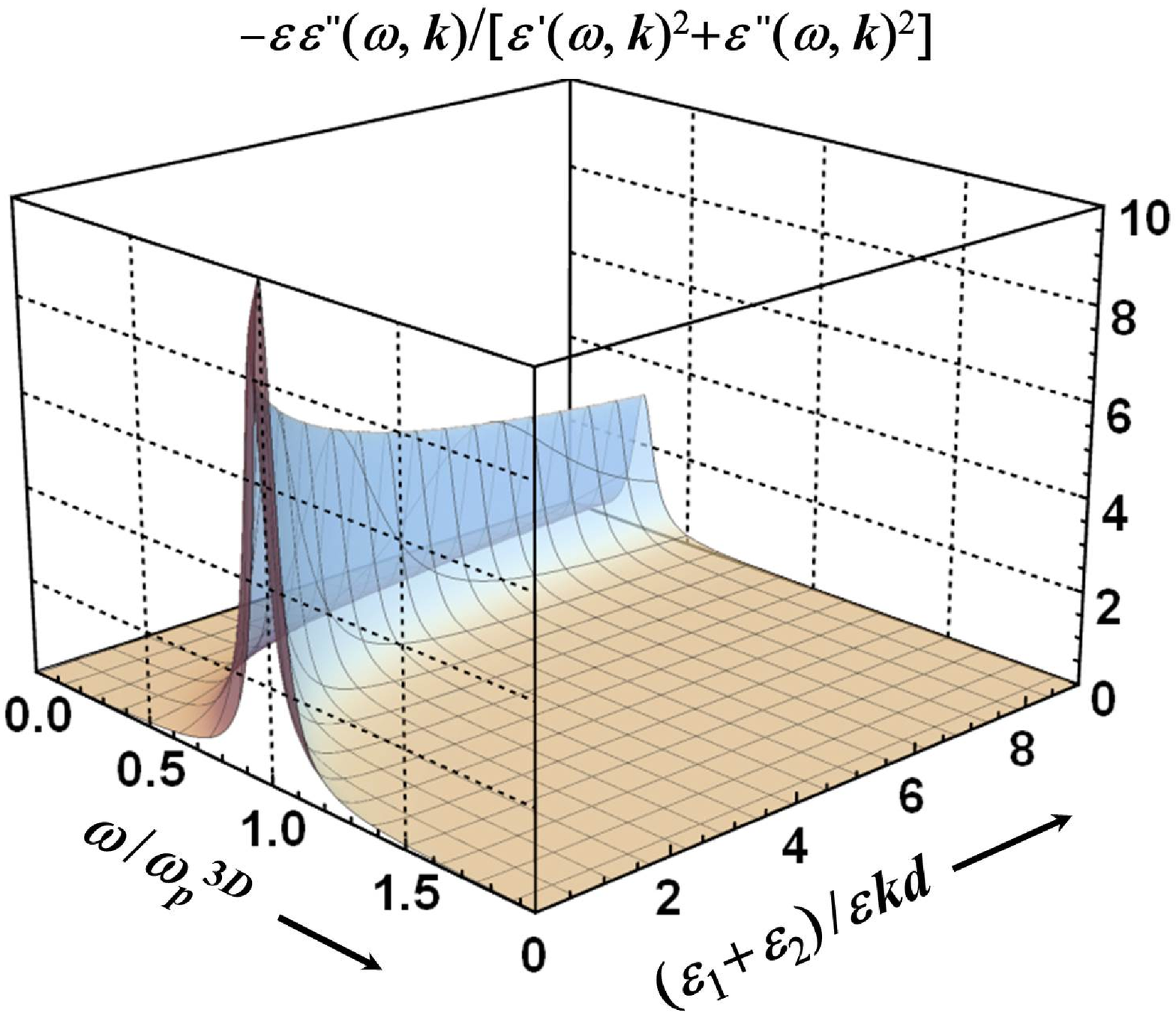}}\caption{(Color online) Plasmon resonance peak behavior given by $-\mbox{Im}[\varepsilon/\varepsilon(k,\omega)]$ of Eq.~(\ref{Lindhard}) as a function of the dimensionless variables $\omega/\omega_p^{3D}$ and $(\varepsilon_1+\varepsilon_2)/\varepsilon kd$.}\label{fig4}
\end{figure}

Expressing all frequency parameters of Eq.~(\ref{Lindhard}) in units of $\omega_p^{3D}$, one obtains the \emph{universal} complex-valued function to feature the dielectric response of the electron gas confined in the finite-thickness plasmonic films. Figures~\ref{fig3}~(a)~and~(b) show its real ($\varepsilon^\prime\!/\varepsilon$) and imaginary ($\varepsilon^{\prime\prime}\!/\varepsilon$) parts as functions of the dimensionless variables $\omega/\omega_p^{3D}$ and $(\varepsilon_1+\varepsilon_2)/\varepsilon kd$, and we also show in Fig.~\ref{fig4} the plasmon peak behavior given by $-\mbox{Im}[\varepsilon/\varepsilon(k,\omega)]$ as a function of the same variables. All graphs are calculated with a moderate parameter ratio $\gamma/\omega_p^{3D}\!=0.1$. In Figure~\ref{fig3}~(b) we see the approach of $\varepsilon^{\prime\prime}\!/\varepsilon$ to the horizontal axis and the shift of the zero point of $\varepsilon^\prime\!/\varepsilon$ from unity at $(\varepsilon_1+\varepsilon_2)/\varepsilon kd\!\ll\!1$ (relatively thick film) towards values lower than unity as $(\varepsilon_1+\varepsilon_2)/\varepsilon kd$ increases to approach the ultrathin film limit at $(\varepsilon_1+\varepsilon_2)/\varepsilon kd\gg\!1$. These correspond to the dissipative loss being decreased at a \emph{fixed} frequency and the plasma frequency being red shifted to go lower than $\omega_p^{3D}$ with the film thickness reduction, which agrees well with the recent measurements done on ultrathin TiN films of controlled variable thickness~\cite{Shalaevgroup17}. At the same time, the red shift of the plasma frequency is accompanied by the gradual increase of the dissipative loss at the \emph{plasma} frequency. This is clearly seen in Fig.~\ref{fig3}~(b) as the $\varepsilon^{\prime\prime}\!/\varepsilon$ magnitude rise in the zeros of the respective $\varepsilon^\prime\!/\varepsilon$ as one moves along the blue arrow, resulting in the plasmon peak red shift and broadening with increasing $(\varepsilon_1+\varepsilon_2)/\varepsilon kd$ as shown in Fig.~\ref{fig4}. We stress that all these features described are universal, peculiar to all plasmonic thin films. Their specific manifestation in real experimental systems is controlled by the film thickness $d$, by the plasma frequency $\omega_p^{3D}$, and by the relative dielectric constant $(\varepsilon_1+\varepsilon_2)/\varepsilon$.

In summary, we predict universal confinement related effects in the optical response of thin plasmonic films as their thickness decreases. While being constant for relatively thick films, the plasma frequency acquires spatial dispersion $\sim\!\sqrt{\varepsilon kd/(\varepsilon_1+\varepsilon_2)}$ typical of 2D materials such as graphene~\cite{Basov}, gradually shifting to the red at all fixed $k$ with the film thickness reduction. The dissipative loss, while decreasing at any fixed frequency, gradually goes up at the plasma frequency as it shifts to the red with the film thickness reduced. These features offer a controllable way to adjust the spatial dispersion and related optical properties of plasmonic thin films and metasurfaces, in particular, not only by varying their material composition~\cite{Shalaevgroup13}, but also by precisely controlling their thickness~\cite{Shalaevgroup17} and by choosing surrounding substrate and superstrate materials appropriately.

We acknowledge fruitful discussions with Alexandra Boltasseva and Harsha Reddy, College of Engineering at Purdue. I.V.B is supported by NSF-ECCS-1306871. V.M.S is supported by NSF-DMR-1506775.

\end{document}